\def\temp{1.34}%
\let\tempp=\relax
\expandafter\ifx\csname psboxversion\endcsname\relax
\else
    \ifdim\temp cm>\psboxversion cm
    \else
      \let\temp=\psboxversion
      \let\tempp= 
    \fi
\fi
\tempp
\let\psboxversion=\temp
\catcode`\@=11
\def\psfortextures{
\def\PSspeci@l##1##2{%
\special{illustration ##1\space scaled ##2}%
}}%
\def\psfordvitops{
\def\PSspeci@l##1##2{%
\special{dvitops: import ##1\space \the\drawingwd \the\drawinght}%
}}%
\def\psfordvips{
\def\PSspeci@l##1##2{%
\d@my=0.1bp \d@mx=\drawingwd \divide\d@mx by\d@my
\includegraphics{##1\space}}}%
\def\psforoztex{
\def\PSspeci@l##1##2{%
\special{##1 \space
      ##2 1000 div dup scale
      \number-\psllx\space \number-\pslly\space translate
}}}%
\def\psfordvitps{
\def\psdimt@n@sp##1{\d@mx=##1\relax\edef\psn@sp{\number\d@mx}}
\def\PSspeci@l##1##2{%
\special{dvitps: Include0 "psfig.psr"}
\psdimt@n@sp{\drawingwd}
\special{dvitps: Literal "\psn@sp\space"}
\psdimt@n@sp{\drawinght}
\special{dvitps: Literal "\psn@sp\space"}
\psdimt@n@sp{\psllx bp}
\special{dvitps: Literal "\psn@sp\space"}
\psdimt@n@sp{\pslly bp}
\special{dvitps: Literal "\psn@sp\space"}
\psdimt@n@sp{\psurx bp}
\special{dvitps: Literal "\psn@sp\space"}
\psdimt@n@sp{\psury bp}
\special{dvitps: Literal "\psn@sp\space startTexFig\space"}
\special{dvitps: Include1 "##1"}
\special{dvitps: Literal "endTexFig\space"}
}}%
\def\psfordvialw{
\def\PSspeci@l##1##2{
\special{language "PostScript",
position = "bottom left",
literal "  \psllx\space \pslly\space translate
  ##2 1000 div dup scale
  -\psllx\space -\pslly\space translate",
include "##1"}
}}%
\def\psforptips{
\def\PSspeci@l##1##2{{
\d@mx=\psurx bp
\advance \d@mx by -\psllx bp
\divide \d@mx by 1000\multiply\d@mx by \xscale
\incm{\d@mx}
\let\tmpx\dimincm
\d@my=\psury bp
\advance \d@my by -\pslly bp
\divide \d@my by 1000\multiply\d@my by \xscale
\incm{\d@my}
\let\tmpy\dimincm
\d@mx=-\psllx bp
\divide \d@mx by 1000\multiply\d@mx by \xscale
\d@my=-\pslly bp
\divide \d@my by 1000\multiply\d@my by \xscale
\at(\d@mx;\d@my){\special{ps:##1 x=\tmpx, y=\tmpy}}
}}}%
\def\psonlyboxes{
\def\PSspeci@l##1##2{%
\at(0cm;0cm){\boxit{\vbox to\drawinght
  {\vss\hbox to\drawingwd{\at(0cm;0cm){\hbox{({\tt##1})}}\hss}}}}
}}%
\def\psloc@lerr#1{%
\let\savedPSspeci@l=\PSspeci@l%
\def\PSspeci@l##1##2{%
\at(0cm;0cm){\boxit{\vbox to\drawinght
  {\vss\hbox to\drawingwd{\at(0cm;0cm){\hbox{({\tt##1}) #1}}\hss}}}}
\let\PSspeci@l=\savedPSspeci@l
}}%
\newread\pst@mpin
\newdimen\drawinght\newdimen\drawingwd
\newdimen\psxoffset\newdimen\psyoffset
\newbox\drawingBox
\newcount\xscale \newcount\yscale \newdimen\pscm\pscm=1cm
\newdimen\d@mx \newdimen\d@my
\newdimen\pswdincr \newdimen\pshtincr
\let\ps@nnotation=\relax
{\catcode`\|=0 |catcode`|\=12 |catcode`|
|catcode`#=12 |catcode`*=14
|xdef|backslashother{\}*
|xdef|percentother{
|xdef|tildeother{~}*
|xdef|sharpother{#}*
}%
\def\R@moveMeaningHeader#1:->{}%
\def\uncatcode#1{%
\edef#1{\expandafter\R@moveMeaningHeader\meaning#1}}%
\def\execute#1{#1}
\def\psm@keother#1{\catcode`#112\relax}
\def\executeinspecs#1{%
\execute{\begingroup\let\do\psm@keother\dospecials\catcode`\^^M=9#1\endgroup}}%
\def\@mpty{}%
\def\matchexpin#1#2{
  \fi%
  \edef\tmpb{{#2}}%
  \expandafter\makem@tchtmp\tmpb%
  \edef\tmpa{#1}\edef\tmpb{#2}%
  \expandafter\expandafter\expandafter\m@tchtmp\expandafter\tmpa\tmpb\endm@tch%
  \if\match%
}%
\def\matchin#1#2{%
  \fi%
  \makem@tchtmp{#2}%
  \m@tchtmp#1#2\endm@tch%
  \if\match%
}%
\def\makem@tchtmp#1{\def\m@tchtmp##1#1##2\endm@tch{%
  \def\tmpa{##1}\def\tmpb{##2}\let\m@tchtmp=\relax%
  \ifx\tmpb\@mpty\def\match{YN}%
  \else\def\match{YY}\fi%
}}%
\def\incm#1{{\psxoffset=1cm\d@my=#1
 \d@mx=\d@my
  \divide\d@mx by \psxoffset
  \xdef\dimincm{\number\d@mx.}
  \advance\d@my by -\number\d@mx cm
  \multiply\d@my by 100
 \d@mx=\d@my
  \divide\d@mx by \psxoffset
  \edef\dimincm{\dimincm\number\d@mx}
  \advance\d@my by -\number\d@mx cm
  \multiply\d@my by 100
 \d@mx=\d@my
  \divide\d@mx by \psxoffset
  \xdef\dimincm{\dimincm\number\d@mx}
}}%
\newif\ifNotB@undingBox
\newhelp\PShelp{Proceed: you'll have a 5cm square blank box instead of
your graphics (Jean Orloff).}%
\def\s@tsize#1 #2 #3 #4\@ndsize{
  \def\psllx{#1}\def\pslly{#2}%
  \def\psurx{#3}\def\psury{#4}
  \ifx\psurx\@mpty\NotB@undingBoxtrue
  \else
    \drawinght=#4bp\advance\drawinght by-#2bp
    \drawingwd=#3bp\advance\drawingwd by-#1bp
  \fi
  }%
\def\sc@nBBline#1:#2\@ndBBline{\edef\p@rameter{#1}\edef\v@lue{#2}}%
\def\g@bblefirstblank#1#2:{\ifx#1 \else#1\fi#2}%
{\catcode`\%=12
\xdef\B@undingBox{
\def\ReadPSize#1{
 \readfilename#1\relax
 \let\PSfilename=\lastreadfilename
 \openin\pst@mpin=#1\relax
 \ifeof\pst@mpin \errhelp=\PShelp
   \errmessage{I haven't found your postscript file (\PSfilename)}%
   \psloc@lerr{was not found}%
   \s@tsize 0 0 142 142\@ndsize
   \closein\pst@mpin
 \else
   \if\matchexpin{\GlobalInputList}{, \lastreadfilename}%
   \else\xdef\GlobalInputList{\GlobalInputList, \lastreadfilename}%
     \immediate\write\psbj@inaux{\lastreadfilename,}%
   \fi%
   \loop
     \executeinspecs{\catcode`\ =10\global\read\pst@mpin to\n@xtline}%
     \ifeof\pst@mpin
       \errhelp=\PShelp
       \errmessage{(\PSfilename) is not an Encapsulated PostScript File:
           I could not find any \B@undingBox: line.}%
       \edef\v@lue{0 0 142 142:}%
       \psloc@lerr{is not an EPSFile}%
       \NotB@undingBoxfalse
     \else
       \expandafter\sc@nBBline\n@xtline:\@ndBBline
       \ifx\p@rameter\B@undingBox\NotB@undingBoxfalse
         \edef\t@mp{%
           \expandafter\g@bblefirstblank\v@lue\space\space\space}%
         \expandafter\s@tsize\t@mp\@ndsize
       \else\NotB@undingBoxtrue
       \fi
     \fi
   \ifNotB@undingBox\repeat
   \closein\pst@mpin
 \fi
\message{#1}%
}%
\def\psboxto(#1;#2)#3{\vbox{%
   \ReadPSize{#3}%
   \advance\pswdincr by \drawingwd
   \advance\pshtincr by \drawinght
   \divide\pswdincr by 1000
   \divide\pshtincr by 1000
   \d@mx=#1
   \ifdim\d@mx=0pt\xscale=1000
         \else \xscale=\d@mx \divide \xscale by \pswdincr\fi
   \d@my=#2
   \ifdim\d@my=0pt\yscale=1000
         \else \yscale=\d@my \divide \yscale by \pshtincr\fi
   \ifnum\yscale=1000
         \else\ifnum\xscale=1000\xscale=\yscale
                    \else\ifnum\yscale<\xscale\xscale=\yscale\fi
              \fi
   \fi
   \divide\drawingwd by1000 \multiply\drawingwd by\xscale
   \divide\drawinght by1000 \multiply\drawinght by\xscale
   \divide\psxoffset by1000 \multiply\psxoffset by\xscale
   \divide\psyoffset by1000 \multiply\psyoffset by\xscale
   \global\divide\pscm by 1000
   \global\multiply\pscm by\xscale
   \multiply\pswdincr by\xscale \multiply\pshtincr by\xscale
   \ifdim\d@mx=0pt\d@mx=\pswdincr\fi
   \ifdim\d@my=0pt\d@my=\pshtincr\fi
   \message{scaled \the\xscale}%
 \hbox to\d@mx{\hss\vbox to\d@my{\vss
   \global\setbox\drawingBox=\hbox to 0pt{\kern\psxoffset\vbox to 0pt{%
      \kern-\psyoffset
      \PSspeci@l{\PSfilename}{\the\xscale}%
      \vss}\hss\ps@nnotation}%
   \global\wd\drawingBox=\the\pswdincr
   \global\ht\drawingBox=\the\pshtincr
   \global\drawingwd=\pswdincr
   \global\drawinght=\pshtincr
   \baselineskip=0pt
   \copy\drawingBox
 \vss}\hss}%
  \global\psxoffset=0pt
  \global\psyoffset=0pt
  \global\pswdincr=0pt
  \global\pshtincr=0pt 
  \global\pscm=1cm 
}}%
\def\psboxscaled#1#2{\vbox{%
  \ReadPSize{#2}%
  \xscale=#1
  \message{scaled \the\xscale}%
  \divide\pswdincr by 1000 \multiply\pswdincr by \xscale
  \divide\pshtincr by 1000 \multiply\pshtincr by \xscale
  \divide\psxoffset by1000 \multiply\psxoffset by\xscale
  \divide\psyoffset by1000 \multiply\psyoffset by\xscale
  \divide\drawingwd by1000 \multiply\drawingwd by\xscale
  \divide\drawinght by1000 \multiply\drawinght by\xscale
  \global\divide\pscm by 1000
  \global\multiply\pscm by\xscale
  \global\setbox\drawingBox=\hbox to 0pt{\kern\psxoffset\vbox to 0pt{%
     \kern-\psyoffset
     \PSspeci@l{\PSfilename}{\the\xscale}%
     \vss}\hss\ps@nnotation}%
  \advance\pswdincr by \drawingwd
  \advance\pshtincr by \drawinght
  \global\wd\drawingBox=\the\pswdincr
  \global\ht\drawingBox=\the\pshtincr
  \global\drawingwd=\pswdincr
  \global\drawinght=\pshtincr
  \baselineskip=0pt
  \copy\drawingBox
  \global\psxoffset=0pt
  \global\psyoffset=0pt
  \global\pswdincr=0pt
  \global\pshtincr=0pt 
  \global\pscm=1cm
}}%
\def\psbox#1{\psboxscaled{1000}{#1}}%
\newif\ifn@teof\n@teoftrue
\newif\ifc@ntrolline
\newif\ifmatch
\newread\j@insplitin
\newwrite\j@insplitout
\newwrite\psbj@inaux
\immediate\openout\psbj@inaux=psbjoin.aux
\immediate\write\psbj@inaux{\string\joinfiles}%
\immediate\write\psbj@inaux{\jobname,}%
\def\toother#1{\ifcat\relax#1\else\expandafter%
  \toother@ux\meaning#1\endtoother@ux\fi}%
\def\toother@ux#1 #2#3\endtoother@ux{\def\tmp{#3}%
  \ifx\tmp\@mpty\def\tmp{#2}\let\next=\relax%
  \else\def\next{\toother@ux#2#3\endtoother@ux}\fi%
\next}%
\let\readfilenamehook=\relax
\def\re@d{\expandafter\re@daux}
\def\re@daux{\futurelet\nextchar\stopre@dtest}%
\def\re@dnext{\xdef\lastreadfilename{\lastreadfilename\nextchar}%
  \afterassignment\re@d\let\nextchar}%
\def\stopre@d{\egroup\readfilenamehook}%
\def\stopre@dtest{%
  \ifcat\nextchar\relax\let\nextread\stopre@d
  \else
    \ifcat\nextchar\space\def\nextread{%
      \afterassignment\stopre@d\chardef\nextchar=`}%
    \else\let\nextread=\re@dnext
      \toother\nextchar
      \edef\nextchar{\tmp}%
    \fi
  \fi\nextread}%
\def\readfilename{\bgroup%
  \let\\=\backslashother \let\%=\percentother \let\~=\tildeother
  \let\#=\sharpother \xdef\lastreadfilename{}%
  \re@d}%
\xdef\GlobalInputList{\jobname}%
\def\psnewinput{%
  \def\readfilenamehook{
    \if\matchexpin{\GlobalInputList}{, \lastreadfilename}%
    \else\xdef\GlobalInputList{\GlobalInputList, \lastreadfilename}%
      \immediate\write\psbj@inaux{\lastreadfilename,}%
    \fi%
    \ps@ldinput\lastreadfilename\relax%
    \let\readfilenamehook=\relax%
  }\readfilename%
}%
\expandafter\ifx\csname @@input\endcsname\relax    
  \immediate\let\ps@ldinput=\input\def\input{\psnewinput}%
\else
  \immediate\let\ps@ldinput=\@@input
  \def\@@input{\psnewinput}%
\fi%
\def\nowarnopenout{%
 \def\warnopenout##1##2{%
   \readfilename##2\relax
   \message{\lastreadfilename}%
   \immediate\openout##1=\lastreadfilename\relax}}%
\def\warnopenout#1#2{%
 \readfilename#2\relax
 \def\t@mp{TrashMe,psbjoin.aux,psbjoint.tex,}\uncatcode\t@mp
 \if\matchexpin{\t@mp}{\lastreadfilename,}%
 \else
   \immediate\openin\pst@mpin=\lastreadfilename\relax
   \ifeof\pst@mpin
     \else
     \errhelp{If the content of this file is so precious to you, abort (ie
press x or e) and rename it before retrying.}%
     \errmessage{I'm just about to replace your file named \lastreadfilename}%
   \fi
   \immediate\closein\pst@mpin
 \fi
 \message{\lastreadfilename}%
 \immediate\openout#1=\lastreadfilename\relax}%
{\catcode`\%=12\catcode`\*=14
\gdef\splitfile#1{*
 \readfilename#1\relax
 \immediate\openin\j@insplitin=\lastreadfilename\relax
 \ifeof\j@insplitin
   \message{! I couldn't find and split \lastreadfilename!}*
 \else
   \immediate\openout\j@insplitout=TrashMe
   \message{< Splitting \lastreadfilename\space into}*
   \loop
     \ifeof\j@insplitin
       \immediate\closein\j@insplitin\n@teoffalse
     \else
       \n@teoftrue
       \executeinspecs{\global\read\j@insplitin to\spl@tinline\expandafter
         \ch@ckbeginnewfile\spl@tinline
       \ifc@ntrolline
       \else
         \toks0=\expandafter{\spl@tinline}*
         \immediate\write\j@insplitout{\the\toks0}*
       \fi
     \fi
   \ifn@teof\repeat
   \immediate\closeout\j@insplitout
 \fi\message{>}*
}*
\gdef\ch@ckbeginnewfile#1
 \def\t@mp{#1}*
 \ifx\@mpty\t@mp
   \def\t@mp{#3}*
   \ifx\@mpty\t@mp
     \global\c@ntrollinefalse
   \else
     \immediate\closeout\j@insplitout
     \warnopenout\j@insplitout{#2}*
     \global\c@ntrollinetrue
   \fi
 \else
   \global\c@ntrollinefalse
 \fi}*
\gdef\joinfiles#1\into#2{*
 \message{< Joining following files into}*
 \warnopenout\j@insplitout{#2}*
 \message{:}*
 {*
 \edef\w@##1{\immediate\write\j@insplitout{##1}}*
\w@{
\w@{
\w@{
\w@{
\w@{
\w@{
\w@{
\w@{
\w@{
\w@{
\w@{\string\input\space psbox.tex}*
\w@{\string\splitfile{\string\jobname}}*
\w@{\string\let\string\autojoin=\string\relax}*
}*
 \expandafter\tre@tfilelist#1, \endtre@t
 \immediate\closeout\j@insplitout
 \message{>}*
}*
\gdef\tre@tfilelist#1, #2\endtre@t{*
 \readfilename#1\relax
 \ifx\@mpty\lastreadfilename
 \else
   \immediate\openin\j@insplitin=\lastreadfilename\relax
   \ifeof\j@insplitin
     \errmessage{I couldn't find file \lastreadfilename}*
   \else
     \message{\lastreadfilename}*
     \immediate\write\j@insplitout{
     \executeinspecs{\global\read\j@insplitin to\oldj@ininline}*
     \loop
       \ifeof\j@insplitin\immediate\closein\j@insplitin\n@teoffalse
       \else\n@teoftrue
         \executeinspecs{\global\read\j@insplitin to\j@ininline}*
         \toks0=\expandafter{\oldj@ininline}*
         \let\oldj@ininline=\j@ininline
         \immediate\write\j@insplitout{\the\toks0}*
       \fi
     \ifn@teof
     \repeat
   \immediate\closein\j@insplitin
   \fi
   \tre@tfilelist#2, \endtre@t
 \fi}*
}%
\def\autojoin{%
 \immediate\write\psbj@inaux{\string\into{psbjoint.tex}}%
 \immediate\closeout\psbj@inaux
 \expandafter\joinfiles\GlobalInputList\into{psbjoint.tex}%
}%
\def\centinsert#1{\midinsert\line{\hss#1\hss}\endinsert}%
\def\psannotate#1#2{\vbox{%
  \def\ps@nnotation{#2\global\let\ps@nnotation=\relax}#1}}%
\def\pscaption#1#2{\vbox{%
   \setbox\drawingBox=#1
   \copy\drawingBox
   \vskip\baselineskip
   \vbox{\hsize=\wd\drawingBox\setbox0=\hbox{#2}%
     \ifdim\wd0>\hsize
       \noindent\unhbox0\tolerance=5000
    \else\centerline{\box0}%
    \fi
}}}%
\def\at(#1;#2)#3{\setbox0=\hbox{#3}\ht0=0pt\dp0=0pt
  \rlap{\kern#1\vbox to0pt{\kern-#2\box0\vss}}}%
\newdimen\gridht \newdimen\gridwd
\def\gridfill(#1;#2){%
  \setbox0=\hbox to 1\pscm
  {\vrule height1\pscm width.4pt\leaders\hrule\hfill}%
  \gridht=#1
  \divide\gridht by \ht0
  \multiply\gridht by \ht0
  \gridwd=#2
  \divide\gridwd by \wd0
  \multiply\gridwd by \wd0
  \advance \gridwd by \wd0
  \vbox to \gridht{\leaders\hbox to\gridwd{\leaders\box0\hfill}\vfill}}%
\def\fillinggrid{\at(0cm;0cm){\vbox{%
  \gridfill(\drawinght;\drawingwd)}}}%
\def\textleftof#1:{%
  \setbox1=#1
  \setbox0=\vbox\bgroup
    \advance\hsize by -\wd1 \advance\hsize by -2em}%
\def\textrightof#1:{%
  \setbox0=#1
  \setbox1=\vbox\bgroup
    \advance\hsize by -\wd0 \advance\hsize by -2em}%
\def\endtext{%
  \egroup
  \hbox to \hsize{\valign{\vfil##\vfil\cr%
\box0\cr%
\noalign{\hss}\box1\cr}}}%
\def\frameit#1#2#3{\hbox{\vrule width#1\vbox{%
  \hrule height#1\vskip#2\hbox{\hskip#2\vbox{#3}\hskip#2}%
        \vskip#2\hrule height#1}\vrule width#1}}%
\def\boxit#1{\frameit{0.4pt}{0pt}{#1}}%
\catcode`\@=12 
 \psfordvips   

\documentstyle[referee]{l-aa}
\begin{document}
\newcommand{\be}{\begin{eqnarray}}
\newcommand{\ee}{\end{eqnarray}}
\def\a{\"a}
\def\A{\"A}
\def\o{\"o}
\def\O{\"O}
\thesaurus{01(12.04.3; 08.22.1; 03.13.6; 11.04.1; 11.01.1)}

\title{Cepheid metallicity and Hubble constant}

\author{J. Nevalainen \inst{1}
\and M. Roos \inst{2} }

\institute{Harvard Smithsonian Center for Astrophysics, 60 Garden
Street, Cambridge, MA 01238, USA and \\
Helsinki University Observatory, P.O.B. 14,
FIN--00014 UNIVERSITY OF HELSINKI, Finland
\and  Physics Department, High Energy Physics Division,
P.O.B. 9, FIN--00014 UNIVERSITY OF HELSINKI, Finland}

\offprints{J. Nevalainen, e-mail: jukka@head-cfa.harvard.edu}

\date{Received  ; accepted }

\maketitle

\begin{abstract} 

This study combines data from a set of seven Cepheid-calibrated galaxies
in Virgo, Leo I and Fornax and four supernov\ae\ with known
metallicities, including metallicity corrections to the Cepheid
distances, in order to determine a consistent mean value for the
Hubble constant.  We derive Virgo and Coma distances over different
paths, starting from the metallicity corrected Cepheid distances of
Virgo and Leo I galaxies, and we relate them to the Virgo and the Coma
recession velocities, paying special attention to thereby introduced
correlations.  We also determine the velocity of the Fornax cluster and
relate this velocity to its distance.  In addition we use the standard
constant maximum brightness SN Ia method, as well as results from multicolor light
curve shape (MLCS) analysis of SNe Ia, to derive the Hubble constant.  Using
the requirement of statistical consistency, our data set determines the
value of the metallicity coefficient (the change in magnitude per factor
of ten in metal abundance) to be $0.31^{+0.15}_{-0.14}$ in good
agreement with other known determinations.  We also use consistency as a
criterion to select between mutually exclusive data sets, such as the
two known, but conflicting, Virgo recession velocities (in favor of
the lower value), as well as between two different analyses of
(partly) the same supernov\ae\ (in favor of MLCS analysis).  
When combining data we pay attention to the statistically correct handling of
common errors.  Our result is a Hubble constant with the value $H_0 =
68\pm 5$ km s$^{-1}$ Mpc$^{-1}$. 
\end{abstract}

\section{THE METALLICITY DEPENDENCE}

The value of the Hubble constant $H_0$ has been notoriously controversial,
not only when determined by different methods, but even values resulting
from the application of a single method to different objects are often
in glaring conflict. The statistical error is of
course large for small data samples, but in addition most determinations
depend on secondary parameters which are ill known or guessed and which
lack reliable error estimates, systematic errors for short.

In this paper we address ourselves to the problem of determining
$H_0$ using data on galaxies with Cepheid-calibrated distances. Then 
two important sources of systematic errors can be discerned: the
effects of the metallicity of the Cepheid  on the
period-luminosity (PL) relation, and the determination of the distance to
the Large Magellanic Cloud (LMC). 
Our main interest in this paper is to examine the metallicity effect,
and we keep the LMC distance modulus at its classical value 
18.5 $\pm$ 0.1 mag (Madore \& Freedman, 1991).
In Section \ref{disc} we discuss recent developments of determining
LMC distance, and check how these will affect our results.

The Cepheid PL relation has been calibrated in the LMC by Madore \& Freedman 
(1991). 
For quite some time it has been speculated that the Cepheid brightness and
therefore the zero point of the PL relation is affected by the metallicity
of the calibrating Cepheids (for a good review see Sasselov et al.,
1997). Sometimes observations have refuted and sometimes confirmed the
existence of this effect. The idea is that metal richer Cepheids would
be brighter and therefore lie further away than what the standard PL
analysis implies. However, the brighter Cepheids are also thought to
be intrinsically redder, which will complicate the analysis, and make
the change of distance to depend on wavelength, or in practise, the
bandpasses used in the photometric analysis. Throughout this paper we
talk about the metallicity effect to Cepheid distance modulus obtained
using V and I bands, because that is the choice of Hubbe Space Telescope
teams.

Although  the shift in distance depends in theory (Stothers, 1988) on the abundance of heavy
elements, the only actually observable "metallicity" is usually 
the oxygen abundance of surrounding HII regions. Therefore in this paper
we actually address ourselves to the systematic shift due to the [O/H]
dependence of the distance moduli. Assuming that the ratio of oxygen abundance of HII regions to the
corresponding value of LMC, equals the corresponding ratio of heavy
element abundances, we interpret the [O/H] - effect as a metallicity effect.

For the metallicity correction of the distance modulus of a Cepheid
we use a form
\be
\delta\mu = \gamma \ ({\rm [O/H] - [O/H]_{LMC}}).
\ee
$\delta\mu$ is the correction needed to the distance modulus obtained
by standard VI photometry ($\mu_{true} = \mu_{standard \ VI} +
\delta\mu$). [O/H] denotes the logarithm of the abundance of oxygen (by
number) relative to hydrogen, in Solar units.
The logarithmic dependence in the above formula follows Sasselov et al.,
(1997). The constant $\gamma$ gives the correction needed
for the distance modulus in magnitude per factor of ten in metal abundance
(mag/dex).

The early results (Freedman \& Madore, 1990) indicated a small effect, which was later contested by
Gould (1994) who found a much larger $\gamma$.  At that time the color
variation was not accounted for.  More recently, three new
determinations have become available: Beaulieu et al.  (1997) and
Sasselov et al.  (1997) in the EROS project analyzed very large samples
of Cepheids in the LMC and the SMC and published a $\gamma$ value for the Z difference. We use
$\gamma =
0.48^{+0.1}_{-0.2}$ mag/dex 
(Sasselov, priv.comm.) which corresponds to the [O/H] difference between LMC and SMC.
In statistical
agreement with this Kennicutt et al.  (1998) have reported $\gamma =
0.24 \pm 0.16$ mag/dex from a study targeting Cepheids in two fields in
M101.  The third result, $\gamma = 0.4 \pm 0.2$ mag/dex, 
is due to Kochanek (1997) who used multicolor photometry to determine 694
Cepheid distances in 17 galaxies, including the effects of temperature,
extinction and metallicity. 
His simultaneous fitting approach to actual photometry data is
perhaps the currently best way to address the
question of metallicity effects on the cepheid PL relation, but it is rather
complicated (a multi-parameter fitting procedure in a Bayesian approach).
The approach adopted by us is different, we do not consider the
systematic errors of the actual photometry, but instead we put emphasis
on the error correlations of photometry results (magnitudes) 
and on error correlations arising from the metallicity correction
in the averaging procedure.

Beaulieu et al.  (1997) and Sasselov et al.  (1997) applied their
metallicity correction to reevaluate the Hubble constant for Cepheids in
HST-observed galaxies with reported metallicities, and demonstrated
that some of the previous disagreement between the $H_0$ values was
indeed removed.  

In the present paper we investigate whether the inclusion of the
metallicity correction is enough to produce a statistically consistent
$H_0$ average for a larger sample of galaxies, or whether
inconsistencies remain, indicating the presence of yet further
systematic errors.  Our sample consists of 7 galaxies with known
distances, velocities and metallicities, and 4 supernov\ae\ with
known V and B peak absolute magnitudes and metallicities. 

\section{THE DATA SETS}        

The galaxies we have selected for this study fall into five sets due to
differences in the method of analysis.  All have Cepheid-calibrated
distances to which we apply metallicity corrections, and their
metallicities are all known. 

Most of the metallicity values are from Kochanek (1997) taking
the positions of the Cepheids and the metallicity gradient across the
galaxies into account.  We have estimated the metallicities
for NGC 4571 and NGC 1365 from Skillman et al.  (1996) and Zaritsky \&
al.  (1994), respectively.  To all the metallicity values we assign an
error of 0.15 dex, following Kennicutt et al. (1998). This value also
encompasses all the observational [O/H] errors of these galaxies
(Zaritsky et al., 1994). 

The five data sets (with our code names) are briefly summarized below;
more detail can be found in Table 1 and in Section 4.

{\sl Virgo (V)} Four galaxies in the Virgo cluster, NGC 4321 (M100), NGC
4496A, NGC 4536, and NGC 4571, for which we take the distance
moduli and errors from Freedman (1996) (however, M100 from
Ferrarese et al.  1996). 

{\sl Leo (L)} Two galaxies in the Leo I group, NGC 3368 (M96) and NGC
3351.  We take the distance moduli from Tanvir et al.  (1995) and Graham
et al.  (1997), respectively. 

{\sl Fornax (F)} One galaxy in the Fornax cluster, NGC 1365, for which
the distance modulus (adding the 3 \% uncertainty
of the distance of NGC 1365 from the Fornax center) is given by Freedman
(1996). 

{\sl Sandage (S)} Four galaxies hosting recent Ia type supernov\ae\ (ignoring
historical ones which may be dubious), NGC 4496A (SN 1960F), NGC 4536
(SN 1981B), NGC 4639 (SN 1990N), and NGC 5253 (SN 1972E).  The V and B
peak absolute magnitudes are from Sandage et al.  (1996). 

{\sl Riess (R)} Three of the supernov\ae\ listed above (all except SN
1960F) have been analysed by Riess et al.  (1996) by an empirical method
that uses multicolor light-curve shapes. 

All the distance moduli depend on the LMC modulus which has commonly
been taken to be $18.50 \pm 0.10$ mag. We keep this value, but we discuss the recent
developments of measuring the LMC modulus, and its effect on our results 
in Section \ref{disc}.

In Table 1 we give the distance moduli and metallicities of the
galaxies used as well as the V and B peak absolute magnitudes of the
supernov\ae . The error due to the LMC modulus error has not been 
included in the values tabulated. 

For most of the time we work with the logarithm of the Hubble constant
rather than with $H_0$ itself.  For the galaxies in the sets {\sl V, L,
F} (each galaxy subscripted $g$) $\log H_0$ has the form
\be \log H_{0,g} = \log (zc) - 0.2 (\mu_g + \delta\mu_g) + 5\ , 
\ee 
where $zc$ is the nonrelativistic recession velocity, $\mu_g$ is the distance
modulus, and $\delta\mu_g$ is the metallicity correction to the distance
modulus from Eq.  (1).  In this formula we have suppressed all the
numerous error terms mentioned before. 

The treatment of the supernov\ae\ in set {\sl S} is different. We have
\be \log H_{0,g} = 0.2 [M_V(max)_g - \delta\mu_g] + \alpha_V\ , 
\ee
where $M_V(max)=m-\mu$ is the V peak absolute magnitude, and analogously
for the B data.  The constant $\alpha$ takes the values $\alpha_V =
5.658 \pm 0.011$ and $\alpha_B = 5.637 \pm 0.011$ (Tammann \& Sandage,
1995).  The two errors just happen to be equal, but are uncorrelated. 

The treatment of the supernov\ae\ in set {\sl R} is the following. 
Riess et al. (1996) determine accurate relative distances to 20 SNe Ia
using their MLCS method, tying their distances to the absolute
Cepheid distance scale via SNe 1972E, 1981B and 1990N, leading to
their result H = 64 $\pm 6$ km/s/Mpc. 
We take into account the
metallicity values of these three SNe , correcting the Riess et al. average
distance to them, and thus correcting also their value of the Hubble constant
starting from their value of $\log H_0(\gamma = 0)$, thus
\be 
\log H_0 = \log H_0(\gamma = 0) - 0.2 \langle\delta\mu\rangle\ .
\ee

The supernova data in the sets {\sl S} and {\sl R} are to some extent
overlapping, thus the 
derived $H_0$ values are correlated and should not be
averaged. The methods are, however very different, and so are the
results.
Using our consistency requirement we want to test whether either set can
be combined with sets {\sl V, L}, and {\sl F}.

\begin{table*}[htbp]
\caption[]{Input data}
\begin{flushleft}
\begin{tabular}{llllll}
\hline\noalign{\smallskip}
Galaxy (supernova) & Group & $\langle \mu \rangle$ & $[O/H]-$ & $M_B(max)$& $M_V(max)$\\  
 & & [mag] & $[O/H]_{LMC}$ & [mag] & [mag]\\  
\noalign{\smallskip}
\hline\noalign{\smallskip}
NGC 4321 & Virgo & $31.04\pm 0.17$ & 0.84 & & \\
NGC 4496A (SN 1960F)& Virgo  & $31.13\pm 0.10$ & 0.00
& $-19.52\pm 0.14$ & $-19.61\pm 0.20$ \\
NGC 4571 & Virgo & $30.87\pm 0.15$ & 0.70 & & \\
NGC 4536 (SN 1981B) & Virgo & $31.10\pm 0.13$ & 0.00
& $-19.29\pm 0.13$ & $-19.32\pm 0.12$ \\
NGC 4639 (SN 1990N) & Virgo & & 0.10
& $-19.30\pm 0.23$ & $-19.39\pm 0.23$ \\
NGC 3368 & Leo I & $30.32\pm 0.16$ & 0.69 & & \\
NGC 3351 & Leo I & $30.01\pm 0.19$ & 0.94 & & \\
NGC 1365 & Fornax & $31.32\pm 0.20$ & 0.10 & & \\
NGC 5253 (SN 1972E) & Centaurus &  & -0.25
& $-19.55\pm 0.23$ & $-19.50\pm 0.21$ \\
\noalign{\smallskip}
\hline
\end{tabular}
\label{tab_data}
\end{flushleft}
\end{table*}

\section{THE STRATEGIES}

With the given data sets there are several questions that we can answer. 
For each question a different strategy has to be followed; in
particular, the treatment of common errors is different. 

The first question is obviously what value of the Hubble constant each
data set determines separately, let us call this Strategy I.  If these
values should be final, they must contain all known errors, including the
common ones such as the LMC distance error.  On the other hand, they can
then not be combined into a final average, because that requires that
common errors be included only after averaging.  Thus final averaging
requires a different strategy, let us call that Strategy IV. 

The purpose of Strategy II is to distinguish between conflicting data
sets.  As already mentioned, the correlation between the data
sets {\sl S} and {\sl R} does not permit both of them to be included in
a final average, thus a choice must be made.  Another choice which has
to be made is between the two conflicting values of the Virgo
velocity, in respect to the Local Group,
$1179 \pm 17$ km/s (Sandage \& Tammann 1990) and $1404 \pm 80$ km/s
(Huchra 1988) which cannot be combined.  One of them (at least) must be
wrong. The small error of the low Virgo velocity is probably 
significantly underestimated (Huchra 1997),
but for the time being, due to the lack of a better estimate, 
and since the distace error dominates over the velocity error for Virgo,
we keep the published value.

Before defining Strategy III, let's make a brief digression.
The coefficient $\gamma$ in the metallicity correction term $\delta\mu$
is a quantity common to all data, thus causing correlations between
different galaxies.  To handle this, we treat $\gamma$ (in Strategies I
and II) as a variable rather than as a known input number with
errors.  If we let $\gamma$ take values in the range 0.0 -- 1.0 [mag/dex],
we cover most of the published values.  Thus the Hubble constant is a
function $H_0(\gamma)$.  This permits us to define a Strategy III where
we use the likelihood function formed by all individual galaxies to
"determine" our own best value of $\gamma$.

Clearly, the questions answered by Strategies II and III influence the
way the final average is formed in strategy IV.  There we combine our
$\gamma$ value with the independently determined value of Sasselov et
al. (1997), and Kennicutt et al. (1998), and 
determine the best value of the Hubble constant from the position of
the maximum of the likelihood function, or actually the minimum of the
chi-square sum $d\chi^{2}(\gamma)/d\gamma = 0$.  Finally we add all common errors
neglected up to this point.

\section{CALCULATIONS}
\subsection{Leo-Virgo and Fornax galaxies}

The Hubble constant can be determined from the sets {\sl V} and {\sl L}
by two routes: using the Virgo recession velocity or the Coma recession
velocity.  In addition, the distance to the Virgo center can be
determined directly from Virgo galaxy distances, or from Leo galaxy
distances using published values for the Leo-Virgo distance.  The
distance to Coma can be measured from the above determined Virgo
distance, using published values for the Virgo-Coma distance, or from
the Leo galaxy distances using published values for Leo-Coma distance. 

Making use of all combinations requires a calculational scheme with
quite complicated correlations which we must take into account exactly. 
Errors in common with the distance of the Virgo center are the uncertainty
of the galaxy distances from the Virgo center, and a similar common
error exists for the distance of the Leo center. All individual
determinations of the Virgo distance are affected by the 
uncertainty due to the recession velocity error (as in Freedman et al. 
1994). 

All individual Cepheid distances are further affected by the common LMC
distance error and the errors related to the metallicity correction
term.  As already said, to all the observed metallicity values we
assign, in lack of better information, one and the same error of 0.15
dex, but that does not make it a common error.  The error in $\gamma$ is
indeed common, but we do not introduce it until in Strategy IV. 

The set {\sl F} has its own errors which are not correlated to the
previous ones, except for the LMC distance error and the $\gamma$ error. 
Here the same comment applies as above.  Note in particular that this
set is independent of the conflicting Virgo velocities. 

In order to describe the averaging procedures compactly, we introduce
the notation $\langle A ; B \rangle$ for the weighted mean of two
quantities $A$ and $B$ with errors $\sigma(A)$ and $\sigma(B)$,
variances $V(A) = \sigma(A)^2$ and $V(B) = \sigma(B)^2$, covariance
cov$(A,B)$, and correlation coefficient
corr$(A,B)=$cov$(A,B)/\sigma(A)\sigma(B)$.  For the weighted mean of a
set of quantities $A_i$ we write simply $\langle A\rangle$.  Let us
proceed in numbered steps. 

1) Compute the mean metallicity-corrected distance moduli of the Virgo
galaxies, set {\sl V} 
\be 
\langle\mu_V\rangle = \langle(\mu_g +
\delta\mu_g)_V\rangle 
\ee 
and the mean metallicity-corrected distance
moduli of the Leo galaxies, set {\sl L} 
\be 
\langle\mu_L\rangle =
\langle(\mu_g + \delta\mu_g)_L\rangle\ .  
\ee

The weights are the individual uncorrelated variances $V(\mu_g)$ of the
distance moduli, adding the combined effect of [O/H] - [O/H]$_{LMC}$ 
errors and the $\gamma$ errors in quadrature.  In
Table 1 we only include the $V(\mu_g)$ errors $\sigma(\mu_g)$. 
 
2) Let us denote the Coma velocity $v_C = (7200 \pm 100)$ km/s (Freedman
et al., 1994, and references therein) and the Virgo velocity $v_V$.  In
combining the Virgo and Coma center distances with the Coma and Virgo
velocities, we use both conflicting $v_V$ values, thus we will have two
sets of results: 
high $v_V$ results (denoted by HI) and low $v_V$ results (denoted by LO).
Below we also need the three distance moduli, $\mu_{LV} = 0.99 \pm 0.15$
for the Leo--Virgo distance, $\mu_{LC} = 4.90 \pm 0.34$ for the
Leo--Coma distance, and $\mu_{VC} = 3.75 \pm 0.1$ for the Virgo--Coma
distance (Tanvir et al.  1995 and references therein).  
Since these three values have been determined independently, 
we are not surprised that $\mu_{\rm LC} \ne \mu_{\rm LV} +  \mu_{\rm VC}$,
but we note that both sides of the equation are consistent within
the errors.
Then the logarithm of the combined $H_0$ measurements for the sets {\sl V} and
{\sl L} can be written
\be
\log H_{0,VL} = \langle\log H_{0,V} ; \log H_{0,C}\rangle\ ,
\ee
where
\be
\log H_{0,V} = \log v_V - 0.2 \langle\langle\mu_V\rangle ; 
\langle\mu_L\rangle + \mu_{LV} \rangle + 5\ , 
\ee 
and
\be
\log H_{0,C} = \log v_C - 0.2 \langle\langle\langle\mu_V\rangle ; 
\langle\mu_L\rangle + \mu_{LV}\rangle + \mu_{VC} ; \nonumber \\ 
\langle\mu_L\rangle +  \mu_{LC}\rangle + 5\ .
\ee
Here we do not spell out the various error terms: the LMC magnitude
error $\pm 0.10$, the Virgo back-to-front position error of $\pm 0.35$
mag (Freedman et al.  1994), the distance uncertainty $\pm 0.11$ mag due
to the Virgo recession velocity error, the uncertainty of the Leo galaxy
distances from the center of Leo I group, $\pm 0.04$ mag (Tanvir et al. 
1995), the observational uncertainty of the O-abundance of a HII region
(O/H error), and the uncertainty involved in identifying the heavy
element abundance
of a Cepheid sitting in a HII cloud with the O-abundance of the
surrounding HII cloud (HII-to-ceph error).  At this stage we omit the
last error, we check its effect in the Discussion. 

3) The correlation coefficients needed are found by numerical
integration of the expressions above.  The correlation coefficient
corr$( \langle\langle\langle\mu_V\rangle ; \langle\mu_L\rangle +
\mu_{LV}\rangle + \mu_{VC}$ , $\langle\mu_L\rangle + \mu_{LC}\rangle )$
is of the order of 4--9\% for the metallicity range $\gamma =$ 0.0--1.0. 
The correlation coefficient corr$(\langle\log H_{0,V} , \log
H_{0,C}\rangle\ )$ is of the order of 44--57\% if the LMC error is
neglected, and about 66\% with the LMC error included. 

The combined Virgo-Leo results containing the LMC error 0.10 mag
(Strategy I) are tabulated in Table 2.  Since we do not present results
separately for sets {\sl V} and {\sl L}, we here introduce a different
grouping: the results $H_{0,HI}$ on the first line are obtained by
combining the Coma results with the results using high Virgo velocity;
the results $H_{0,LO}$ on the second line by combining the Coma results
with the results using low Virgo velocity. 

4) The Fornax galaxy NGC 1365 forming the set {\sl F} is by itself
uncorrelated to the previous sets.  Thus the Hubble constant can be
worked out directly.  The Fornax galaxy velocity data (kindly provided
by J.Huchra), consist of 119 galaxies with velocities smaller than 3000
km/s.  The mean velocity and its error from the velocity dispersion are
1459 $\pm$ 35 km/s (consistent with Madore et al. 1998).  
From this error we remove the ``fictitious'' effect
of observational errors of galaxy velocities (Danese et al.  1980),
yielding a true velocity error of 25 km/s.  Correcting for the
Virgo-centric flow with -40 $\pm$ 20 km/s, and converting the velocity
to the Local Group by -147 $\pm$ 10 km/s, gives us the final expansion
velocity of Fornax as 1272 $\pm$ 34 km/s.  
The results for the $\gamma$-dependent Hubble constant $H_{0,F}$ including
the LMC error are tabulated on the third line of Table 2. 

\begin{table*}[htbp]
\caption{$H_0(\gamma)$ for all data sets (H in units of km s$^{-1}$
Mpc$^{-1}$, $\gamma$ in units of mag/dex)}
\begin{flushleft}
\begin{tabular}{lllllll}
\hline\noalign{\smallskip} 
$\gamma $ & 0.0 & 0.2 & 0.4 & 0.6 & 0.8 &1.0\\
\noalign{\smallskip}
\hline\noalign{\smallskip}
$H_{0,HI}$&$80.5\pm5.0$&$77.9\pm 4.9$&$75.2\pm 4.8$&$72.2\pm 4.7$&$69.1\pm 4.6$&$65.9\pm4.6$\\
$H_{0,LO}$&$73.8\pm 4.1$&$71.5\pm 4.0$&$69.1\pm 3.9$&$66.5\pm 3.9$&$63.6\pm 3.9$&$60.6\pm 3.8$\\
$H_{0,F}$ &$69.2\pm 7.4$&$68.6\pm 7.4$&$68.0\pm 7.5$&$67.3\pm 7.7$&$66.7\pm 8.0$&$66.1\pm 8.4$\\
$H_{0,S}$ &$58.3\pm 3.3$&$58.4\pm 3.3$&$58.4\pm 3.4$&$58.5\pm 3.4$&$58.6\pm 3.5$&$58.7\pm 3.7$\\
$H_{0,R}$ &$64.0\pm 4.6$&$64.8\pm 4.7$&$65.5\pm 4.8$&$66.1\pm 5.1$&$66.6\pm 5.3$&$67.0\pm5.6$\\
\noalign{\smallskip}
\hline
\end{tabular}
\end{flushleft}
\label{tab_data}
\end{table*}

\subsection{Supernov\ae }

5) The supernova set {\sl S} is averaged as follows 
\be 
\log H_{0,S} =
\langle 0.2 \langle [M_V(max) - \delta\mu]\rangle + \alpha_V ;\\ 0.2
\langle [M_B(max) - \delta\mu]\rangle + \alpha_B\rangle\ .  
\ee 
The results for the $\gamma$-dependent Hubble constant $H_{0,S}$ including
the LMC error are tabulated on the fourth line of Table 2. 

6) For the supernova set {\sl R} we simply take the published $H_0$
value.  Its error contains the effect of an LMC distance modulus error
0.15 mag, which we decrease to 0.10 mag to be consistent with the
previous part of our analysis, and further we add the metallicity
correction $\langle\delta\mu\rangle$.  The $\gamma$-dependent averages
\be 
\log H_{0,R} = \log H_0 - 0.2 \langle\delta\mu\rangle 
\ee 
are tabulated on the fifth line of Table 2. 

This concludes the calculations in Strategy I, which are summarized as
values of $H_0(\gamma)$ including the LMC error in Table 2.  
By inspection of the results for the different sets one easily convinces
oneself that there is a range of overlap near $\gamma = 0.4-0.8$ for the
sets {\sl LO, F} and {\sl R}.  Does that mean that the two other
results, $H_{0,HI}$ and $H_{0,S}$, can be coldly ignored as being in
statistical disagreement and probably affected by systematic errors? To
answer this question we turn to Strategy II in the next Section.

\section{TESTING}

How important are the disagreements in Table 2? Let us first turn to the
supernova set {\sl S} and test it against the Virgo-Leo sets {\sl HI}
and {\sl LO}.  In this test we do not use the Fornax results, because
{\sl F} is just a single galaxy with a fairly large $H_0$ error, and we
further neglect the set {\sl R} because it anyway does not come into
question to combine it with {\sl S}.  It is important here to remove the
LMC error from the Table 2 results, because it is common to all sets and
completely correlated. 

The first result is that $H_{0,HI}$ disagrees with $H_{0,S}$ by
5.4$\sigma$ at $\gamma=0.0$, diminishing to 3.4$\sigma$ at $\gamma=0.6$. 
Note that no published estimate of $\gamma$ exceeds 0.6 within 1$\sigma$
errors (except Gould 1994, but in that work the color
variation was not accounted for and therefore his $\gamma$ is not
comparable to ours), so it is difficult to defend solutions requiring larger
$\gamma$--values.  Thus we conclude that it is not statistically
defendable to combine $H_{0,HI}$ with $H_{0,S}$. 

The second result is that also $H_{0,LO}$ disagrees with $H_{0,S}$, by
4.9$\sigma$ at $\gamma=0.0$, diminishing to 2.4$\sigma$ at $\gamma=0.6$. 

Since ultimately we have to choose between combining either {\sl S} or
the other supernova set, {\sl R}, with one of the Virgo-Leo sets, let us
see how much better the $H_{0,R}$ fares.  The $H_{0,HI}$ value disagrees
with $H_{0,R}$ by 3.5$\sigma$ at $\gamma=0.0$, but at $\gamma=0.6$ the
disagreement is only 1.3$\sigma$.  The $H_{0,LO}$ value
disagrees with $H_{0,R}$ by 2.4$\sigma$ at $\gamma=0.0$, but from about
$\gamma=0.35$ up the disagreement is no longer significant.
We conclude from this that the set {\sl R} is strongly favored over {\sl
S} on the basis of our consistency requirement. Part of the reason for
this is that the errors of set {\sl S} are very small.

Another test is possible in Strategy III: first we compute for each data
set the mean values of $H_0$, as a function of $\gamma$.  While
averaging $H_{0}$-values at a given $\gamma$-value, we use only
uncorrelated errors as weight, (See Fig. 2 for the {\sl LO + R + F
set}).  
We then construct a $\chi^{2}$ sum as a function of $\gamma$, comparing
the above mean $H_{0}$($\gamma$)-values with the individual 
$H_{0}$($\gamma$)-values from different groups:
each galaxy in the sets
{\sl V, L, F} is represented by one term, the average {\sl S} by one
term, and the average {\sl R} by one term. 

This complicated process of constructing a $\chi^2$ sum from correlated
terms makes it rather problematic to relate it to probability and to
interpret the sum as a goodness-of-fit test.
  However, different data combinations can be compared and
relative figures of merit can be established. 

 \begin{figure*}[htbp]
  \begin{center}
    \leavevmode
    \mbox{\psboxto(0cm;13cm){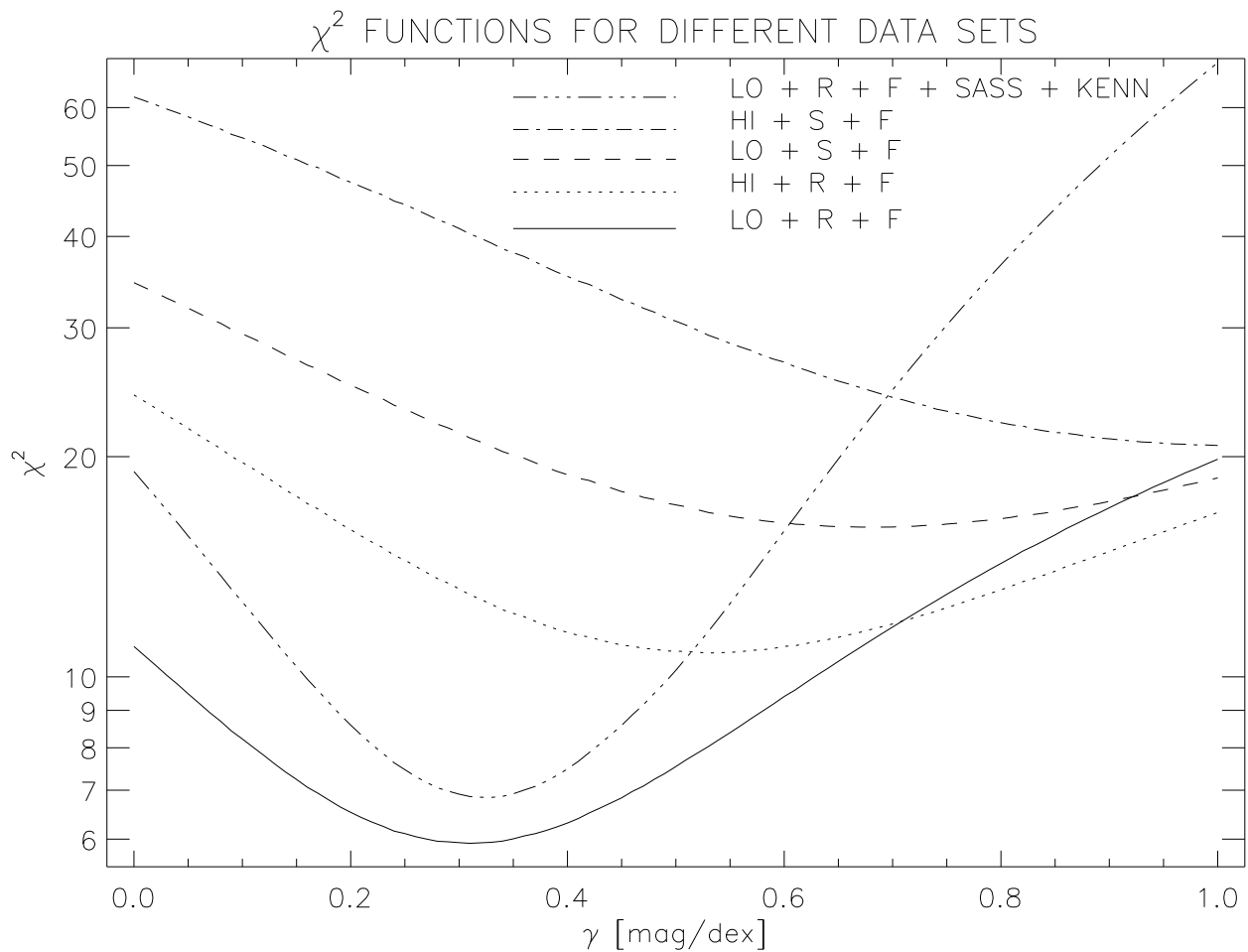}}
   \end{center}
      \caption{The $\chi^2$ sum as a function of $\gamma$ for the four different
      data sets, and for the LO + R + F set including Sasselov's and
      Kennicutt's $\gamma$ as additional terms.} 
           \label{fig_2}
   \end{figure*}

The resulting $\chi^2(\gamma)$ functions are shown in Fig.1 for the four
data set combinations $HI + R + F,\ LO + R + F,\ HI + S + F,\ LO + S +
F$.  The common LMC error is again set to zero.  One notes that the
combination $LO + R + F$ attains the lowest minimum value 5.9. The minima
of all the four combinations of data are
\be
HI + R + F:\ \ \ 10.8\ @\ \gamma=0.53\ , \nonumber\\
LO + R + F:\ \ \ 5.9\ @\ \gamma=0.31\ , \nonumber\\
HI + S + F:\ \ \ \sim 20.7\ @\ \gamma \ge 1.0\ , \nonumber\\
LO + S + F:\ \ \ 16.0\ @\ \gamma=0.69\ . \nonumber
\ee
If we interpret the differences as standard variances,
every combination is at least 2.2$\sigma$ worse than $LO + R + F$.
In particular, every combination including the set $S$ is at least
3.2$\sigma$ worse. From the curves in Fig.1 one can deduce confidence 
intervals for $\gamma$ (cf. Section 6).

An important conclusion is that the overall consistency is worse in the 
combinations including the set $\sl S$ than including the set $\sl R$.

Note that this method also permits us to choose between the high and low Virgo
velocities, although the significance of the test is only 2.2$\sigma$ in
favor of the low Virgo velocity.

 \begin{figure*}[htbp]
  \begin{center}
    \leavevmode
\mbox{\psboxto(0cm;13cm){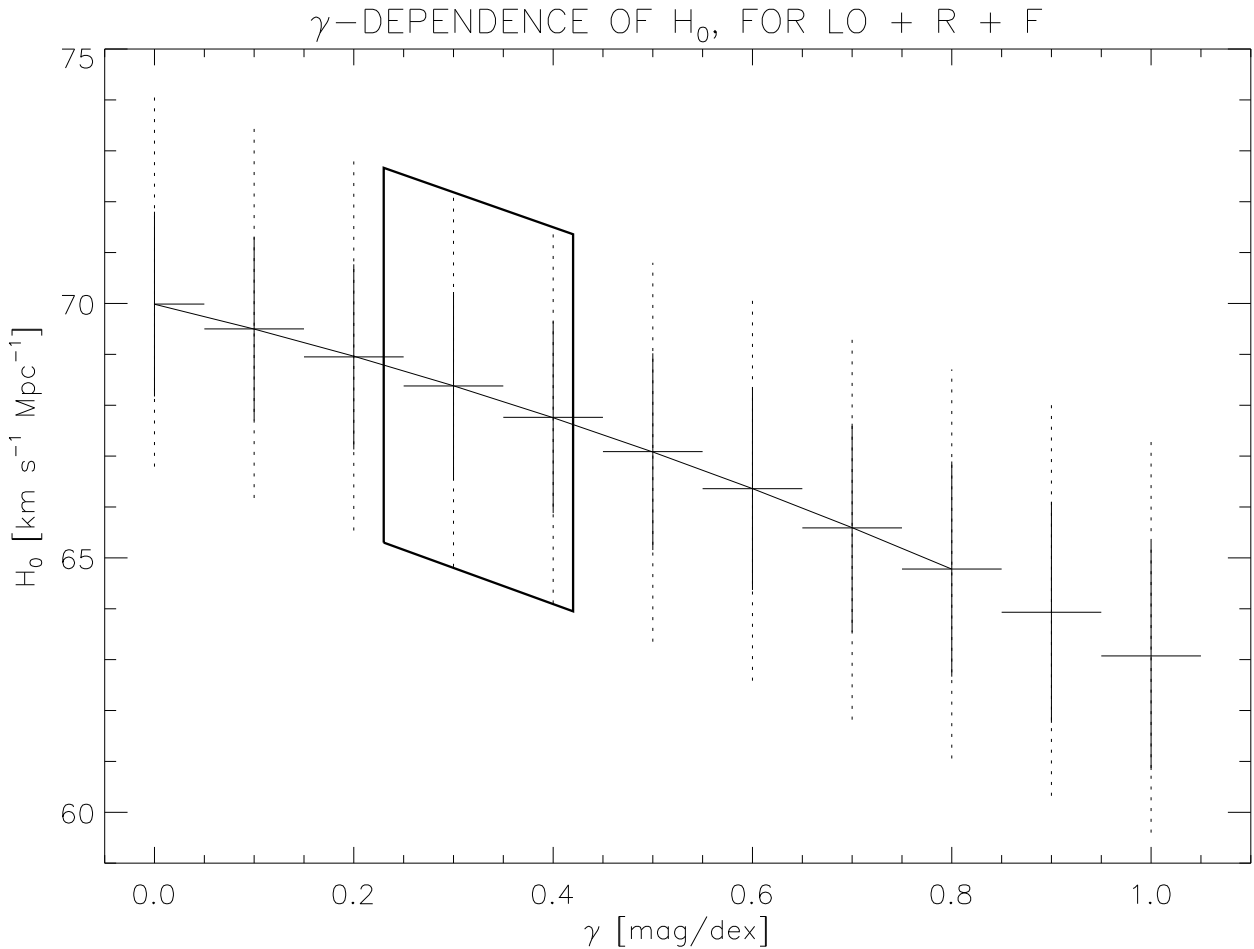}}
  \end{center}
      \caption{The mean $H_{0}$($\gamma$)-values for the data set
      including Leo-Virgo-Coma data using low Virgo velocity, Fornax
      data and SN Ia data from Riess. The errors of the mean (including only
      uncorrelated errors) are indicated with vertical solid lines. 
      The total errors (including LMC error) are
      indicated by vertical dotted lines. The best 2nd order
      polynomial fit is indicated by a solid curve. The $\gamma$-value from our
      analysis, combined with Sasselov's and Kennicutt's value, and the corresponding allowed $H_{0}$ values are
      indicated as a thick quadrangle.}
           \label{fig_1}
   \end{figure*}

\section{RESULTS}

From the $\chi^2$ sum constructed (Strategy III) in the
previous Section and plotted in Fig.1. we can determine a value of $\gamma$.
Keeping both the high and low Virgo velocities, we have two
results. For the set $LO + R + F$ we find
\be 
\gamma=0.31^{+0.15}_{-0.14}\ {\rm mag/dex}\ ,
\ee 
and for $HI + R + F$ 
\be 
\gamma=0.53^{+0.17}_{-0.15}\ {\rm mag/dex}\ .  
\ee 
The low Virgo velocity result is in good agreement with previous results
quoted in Section 1, whereas  the high  Virgo velocity result is only
marginally so. At this point we have to choose between the two.

From our calculations of the mean $H_0(\gamma)$ for {\sl LO + R + F} and
{\sl HI + R + F} we find that the difference $\gamma = 0.53 - 0.31$
corresponds to a shift in $H_0$ of only about 1.4 km s$^{-1}$ Mpc$^{-1}$
which is negligible in comparison with the $H_0$ error flags.  Thus the
disagreement between the two Virgo velocities, which really is of the
order of 2.8$\sigma$ in velocity space, has now shrunk to much less than
one standard deviation in $H_0$ space.  We conclude that we can safely
trust the $\gamma$ value from the low Virgo velocity $\chi^2$ fit which
was favored on the basis of its minimum $\chi^2$ value. 

We can now improve the precision of our $\gamma$ value by combining it
with other independent determinations. The value of Beaulieu et al. (1997)
and Sasselov
et al. (1997) is clearly independent of our result, because it is derived from
Cepheids in the LMC and SMC only. Similarly, 
the Kennicutt et al. (1998) value is derived from observations of M101 only,
and therefore independent of our result.
The Kochanek (1997) value is derived using some of the galaxies we also use, 
thus we cannot consider that completely independent of our analysis.  

Combining our value $\gamma=0.31^{+0.15}_{-0.13}$ mag/dex with the
Sasselov et al. (1997) and Sasselov (priv. comm.) result 
$\gamma = 0.48^{+0.1}_{-0.2}$ mag/dex from
[O/H]$_{\rm SMC}$ / [O/H]$_{\rm LMC}$, and with the Kennicutt et al,
(1998) result 
$\gamma = 0.24 \pm 0.16$ mag/dex,
we obtain the fifth curve in Fig.1 which has its minimum, $\chi^2 = 6.8$ at
\be \gamma=0.33^{+0.09}_{-0.10}\ {\rm mag/dex}\ .  \ee
Note that this implies that the value $\gamma=0$ is rejected by
$3\sigma$.

We finally arrive at Strategy IV.  We transform the above best $\gamma$
value via the fitted function $H_0(\gamma)$ to a $H_0$ value, as in Fig.2.
At this step we have to add the two correlated errors which are
common to all our data sets: the error of 0.10 mag in the LMC distance
modulus, and the above error in $\gamma$.  The effect of the LMC error
we find by redoing all calculations at the fixed LMC distance modulus
values 18.40 and 18.60.  Finally we treat the Virgo velocity error
conflict as a systematic error of size 1.4 km s$^{-1}$ Mpc$^{-1}$ which
we combine in quadrature with the other error terms.  Our result is then
\be
H_0 = 68\pm 5\  {\rm km\ s}^{-1}\ {\rm Mpc}^{-1}\ .
\ee
We note that the dominant component in the error is due to the LMC distance
modulus as long as the HII-to-ceph error is absent (see however Section \ref{disc} for a
discussion of this point).

\section{DISCUSSION}

\label{disc}

To get an idea of the effect of the 
HII-to-ceph error (explained in Section 4), we rerun our computations, using 
an abundance error arbitrarily increased to
0.30 dex for each galaxy, instead of 0.15 dex used
above. The corresponding $\chi^2$-minima become so close to each other that we cannot reject any 
of the data combinations, and therefore we cannot get the best values of $\gamma$ or H$_{0}$.
However, including Sasselov's $\gamma$
value as one term in the $\chi^2$ sum,
we find that the parameters are in the ranges $\gamma$ = 0.25 to 0.63
mag/dex and $H_{0}$ = 58 to 74 km s$^{-1}$ Mpc$^{-1}$. We note
that the error ranges are naturally larger than in our results in the
previous section, but that our value of $H_{0}$ in the previous
section is totally within the new range here.   

There exists quite a range of measured values of the distance modulus
of LMC. 
The Barnes-Evans infrared surface brightness technique to calibrate
period-radius and period-luminosity relations of Galactic Cepheids
lead to a value $\mu_{LMC} = 18.46 \pm 0.02$ (Gieren et al.,
1998), quite close to the classical 18.5 $\pm$ 0.1.

The recent HIPPARCOS measurements of the Galactic Cepheids
lead to a value $\mu_{LMC} = 18.7 \pm 0.1$ (Feast \& Catchpole, 1997).
The reanalysis of the same data gives lower values
$\mu_{LMC} = 18.57 \pm 0.11$ (Madore \& Freedman, 1997) and
$\mu_{LMC} = 18.56 \pm 0.08$ (Oudmaijer et al., 1998).
The most recent determination from the expanding ring around the SN
1987a is $\mu_{LMC} = 18.58 \pm 0.03$ (Panagia et al., 1997).
HIPPARCOS parallax analyses of local sub\-dwarfs lead to values
$\mu_{LMC} = 18.61 \pm 0.07$ (Gratton et al, 1997),
or $\mu_{LMC} = 18.65 \pm 0.10$ (Reid, 1997),
who speculates that the RR Lyrae distance scale is too small.
These observations form a quite impressive body of evidence for the real LMC distance modulus to be
slightly higher than the standard 18.5 mag that we have assumed in this work.

However, the situation is more complicated due
to observations of low values of $\mu_{LMC}$:
calibrating Galactic RR Lyrae variables using HIPPARCOS data
leads to $\mu_{LMC} = 18.37 \pm 0.23$ (Luri et al., 1998) and  
$\mu_{LMC} = 18.26 \pm 0.15$ (Fernley et al, 1998). 
The HIPPARCOS calibrated red clump star method (Paczy\'nski \& Stanek,
1998), leads to low values
$\mu_{LMC} = 18.07 \pm 0.10$ (Stanek et al, 1998)
and
$\mu_{LMC} = 18.08 \pm 0.12$ (Udalski et al., 1998).

It is obvious that all these observations are not consistent with each
other, so 
we may be obliged to make a selection. One
possible way out would be to rely on the revised red clump values
$\mu_{LMC} = 18.28 \pm 0.18$ (Girardi et al, 1998) 
who take metallicity effect into account, and
$\mu_{LMC} = 18.36 \pm 0.17$ (Cole, 1998),
who considers a luminosity dependence of the red clump stars on age and metallicity.
Our motivation is that the red clump method is relatively new, and as
such prone to more systematic errors than other methods. 
One point worth mentioning is that the weighted mean of the LMC modulus
remains robustly 18.50 to within better than 0.02 
regardless of whether we use the published variances as weights or an
arbitrarily chosen weight, common to all data. Also, the mean is 
unsensitive to whether we include or exclude Feast \& Catchpole (1997),
which have already been revised by Stanek et al. (1998) 
and Udalski et al. (1998). 
Taking all the above into account (with only the revised data) we propose
a very conservative estimate for the LMC distance modulus of
18.50$^{+0.15}_{-0.25}$. This confidence interval is chosen to cover all the above quoted
central values. We note that the data set used here is not totally
consistent, and therefore this estimate for the error range is an overestimate, containing
some underlying systematic errors that have not been estimated properly. 
 
Another attempt to reduce some of this bias is by ``weighing''  the
data sets with the merits of the methods used to obtain them. 
Considering the strong evidence for high distance an argument in
disfavor of the red clump method, we are left with three independent
methods (HIPPARCOS cepheids, local subdwarf parallaxes and SN
1987a) giving an LMC distance exceeding 18.50, one giving an
intermediate value (IR surface brightness) and only one (RR Lyrae
method) giving a low distance. This would point to a significant systematic error in RR
Lyrae method, (see Reid's (1997) discussion on this). Excluding RR
Lyrae results we end up with our ``less conservative estimate'', a slightly higher value of about 
$\mu_{LMC} = 18.55 \pm 0.1$.

We conclude that the issue of the LMC distance is not settled yet.
Our two estimates above, a conservative one and a less conservative one
are two different ways to look at the same problem. We note that both distance estimates are consistent with
the classical value of Freedman \& Madore.
In the following we check how our best value would be affected by
adopting either one of the above two ``conservatisms''.

Recalculating with the conservative estimate $\mu_{LMC} = 18.50^{+0.15}_{-0.25}$ we note that
$\gamma$ - determination and therefore the best value of
H$_{0}$ remains unchanged, since the central value of this estimate
equals the classical one, and since LMC distance errors are not included
in $\chi^{2}$-calculations. However, the errors of the final value
increase so that we obtain 
$H_0 = 68^{+8}_{-6}$ km s$^{-1}$ Mpc$^{-1}$. 
Since this error range is conservative and an overestimate, and since
our best value is totally in this range,
this value should only be taken as and indicator that our best value
is valid.

Recalculating with the less conservative estimate $18.55 \pm 0.1$ we find out that all our
conclusions remain unchanged, the values of $\gamma$ remain virtually
unchanged and the Hubble constant decreases marginally, by 2\%, or 1.5 km s$^{-1}$
Mpc$^{-1}$. This is our suggestion for the systematic error of $H_0$
due to the LMC distance uncertainty. However, adding this value in
quadrature to the errors of our best estimate has no effect, therefore
also this estimate supports our best value.

\section{CONCLUSIONS}
\label{conc}

The existence of well measured distances to galaxies for which also the metallicities
are known, permits one to attempt to form a consistent set of data from
which a combined value of the Hubble constant can be obtained. We have
used four Virgo galaxies, two Leo I galaxies, one galaxy in Fornax,
and two partially overlapping analyses of four supernov\ae\ 
(all listed in Section 2. and in Table 1)
to study the metallicity dependence of the Hubble constant derived from these data.

The analysis introduces several types of correlations, and some of the
input errors are correlated as well.  We take care of handling all such
correlations statistically correctly.  We conclude that we can select
between the two supernova 
analyses: the MLCS set (Riess et al, 1996) is consistent with all our other 
                 data sets, whereas the standard constant maximum
brightness analysis of type Ia SNe (Sandage et al, 1996) set is not.
We can also select between the two conflicting Virgo velocity values in the literature (in favor of
the low velocity).

Our main results are two (using $\mu_{LMC} = 18.5 \pm 0.1$): 
a new value for the Hubble constant, 
$H_0 = 68\pm 5$ km s$^{-1}$ Mpc$^{-1}$, and a new value for the 
metallicity coefficient 
(the correction for distance modulus in magnitude per factor of ten in metal
abundance) of $0.31^{+0.15}_{-0.14}$ mag/dex, in excellent agreement with 
previous results.
Thus the metallicity effect is clearly significant -- combining our
$\gamma$ value with that of Sasselov et al. (1997) for [O/H], 
and Kennicutt et al. (1997), we obtain
$\gamma=0.33^{+0.09}_{-0.10}$ which is $3\sigma$ away from zero.
Note the the analysis by Kochanek, yielding $\gamma =
0.4 \pm 0.2$, is in excellent agreement with this.

Taking into account the current knowledge of the LMC distance, we see that our best value 
is even more robust (see Section \ref{disc}).

The general agreement of the final data set with our criterion of
statistical consistency permits us to conclude that no further systematic 
errors are needed beyond those accounted for here.

\begin{acknowledgements}The authors wish to thank J.  Huchra,
C.  S.  Kochanek, A.  G.  Riess, D.  D.  Sasselov, and O.  Vilhu for
useful comments, and J. Huchra for providing us with Fornax data. We
also thank the referee for many useful suggestions. JN
thanks Harvard Smithsonian Center for Astrophysics
for the hospitality. JN thanks the Smithsonian Institute for a
Predoctoral Fellowship, and the Finnish Academy for a supplementary
grant.\end{acknowledgements}


\begin{thebibliography}{}


\bibitem[1997]{beau} 
Beaulieu J. P. et al., 1997, A\&A 318, L47

\bibitem[1998]{cole}
Cole A., astro-ph/9804110

\bibitem[1997]{dan} 
Danese L. et al, 1980,  A\&A 82, 322

\bibitem{hipp}
Feast M. W. \& Catchpole R. M., 1997, MNRAS 286, L1 

\bibitem[1990]{fern}
Fernley J. et al, 1998, A \& A, 330, 515-520, 1998

\bibitem{ferr}
Ferrarese L. et al., 1996, Ap. J. 464, 568

\bibitem[1990]{free1}
Freedman W. L. \& Madore B. F., 1990, ApJ 365, 186  

\bibitem[1994]{free}
Freedman W. L. et al., 1994, Nature 371, 757

\bibitem[1996]{free2}
Freedman W. L., 1996, in {\it Critical Dialogues in Cosmology}, p.92,
ed. Neil Turak, World Scientific Publishing. Also astro-ph/9612024

\bibitem[1998]{gier}
Gieren W. P. et al., 1998, ApJ, 496:17-30, 1998

\bibitem[1998]{gira}
Girardi L. et al, 1998, astro-ph/9805127

\bibitem[1994]{gould}
Gould A., 1994, ApJ 426, 542

\bibitem[1997]{grah}
Graham J. A. et al., 1997, ApJ 477, 535

\bibitem[1997]{grat}
Gratton R. G. et al, 1997, astro-ph/9707107.

\bibitem[1988]{huch} 
Huchra J. in van den Bergh S. \& Pritchet C. J.(eds.), 1988, 
PASP Conf. Series 4, 257

\bibitem[1997]{huch2} 
Huchra J., 1997, in {\it The MSSSO Heron Island Workshop on Peculiar Velocities in the Universe},\\
 http://qso.lanl.gov/$\sim$heron/proceedings.html

\bibitem[1998]{kenn}
Kennicutt R. C. et al., 1998, astro-ph/9712055

\bibitem[1997]{koch}
Kochanek C. S., 1997, astro-ph/9703059

\bibitem[1998]{luri}
Luri X. et al., 1998, astro-ph/9805215

\bibitem[1991]{mad} 
Madore B. F. \& Freedman W. L., 1991, PASP, 103, 933

\bibitem[1997]{mad2}
Madore B. F. \& Freedman W. L., 1997, astro-ph/9707091

\bibitem[1998]{mad3}
Madore B. F. et al., 1998, in press

\bibitem[1998]{oud}
Oudmaijer R. D. et al., 1998, astro-ph/9801093

\bibitem[1997]{pan}
Panagia, N., Gilmozzi, R., Kirshner, R.P. 1997, in {\it SN 1987A: Ten Years After},
eds. M.Phillips and N.Suntzeff, ASP Conf. Ser., in press.

\bibitem[1998]{pacz}
Paczy\'nski B. \& Stanek K. Z., 1998, ApJ, 494, L219, 1998

\bibitem[1997]{reid}
Reid, I. N., 1997, astro-ph/9704078

\bibitem[1996]{riess}
Riess A. G., Press W. H., \& Kirshner R. P., 1996, ApJ 473, 88 

\bibitem[1994]{saha} 
Saha A. et al., 1994, ApJ 425, 14

\bibitem[1990]{sand}
Sandage A. \&  Tammann G. A., 1990, ApJ 365, 1

\bibitem[1994]{sand2} 
Sandage A. et al., 1994, ApJ 423, L13

\bibitem[1996]{sand3} 
Sandage A. et al., 1996, ApJ 460, L15

\bibitem[1997]{sass} 
Sasselov D. D. et al., 1997, A\&A 324, 471 

\bibitem[1996]{zar2} 
Skillman E.D. et al., 1996, ApJ 462, 147

\bibitem[1998]{stan}
Stanek K. Z. et al, 1998, astro-ph/9803181

\bibitem[1998]{stot}
Stothers, R. B., 1988, ApJ, 329, 712

\bibitem[1995]{tamm}
Tammann G. A. \& Sandage A., 1995, ApJ 452, 16

\bibitem[1995]{tan} 
Tanvir N. R. et al., 1995, Nature 377, 27

\bibitem[1998]{udal} 
Udalski A. et al., 1998, astro-ph/9803035

\bibitem[1994]{zar1}
Zaritsky D., Kennicutt R. \& Huchra J., 1994, ApJ 420, 87

\end{thebibliography}
\end{document}